\documentclass[Physsubmission, Phys]{SciPost}

\hypersetup{
    colorlinks,
    linkcolor={red!50!black},
    citecolor={blue!50!black},
    urlcolor={blue!80!black}
}

\usepackage[bitstream-charter]{mathdesign}
\usepackage{bm}

\usepackage[draft]{changes}
\definechangesauthor[color=green!50!black]{Wim}
\definechangesauthor[color=red]{Bernard}
\definechangesauthor[color=blue]{Lech}

\DeclareSymbolFont{usualmathcal}{OMS}{cmsy}{m}{n}
\DeclareSymbolFontAlphabet{\mathcal}{usualmathcal}

\begin{document}

\begin{center}{\Large \textbf{
Accessing quark GPDs in diffractive events at an electron-ion collider\\
}}\end{center}

\begin{center}
W. Cosyn\textsuperscript{1,2$\star$},
B. Pire\textsuperscript{3} and
L. Szymanowski\textsuperscript{4}
\end{center}

\begin{center}
{\bf 1}  Department of Physics, Florida International University,  Miami, Florida 3199, USA
\\
{\bf 2} Department of Physics and Astronomy, Ghent University, B9000 Ghent, Belgium
\\
{\bf 3} CPHT, CNRS, \'Ecole Polytechnique, I. P. Paris,
  91128 Palaiseau,     France 
\\
{\bf 4} National Centre for Nuclear Research (NCBJ), Pasteura 7, 02-093 Warsaw, Poland
\\
* wcosyn@fiu.edu
\end{center}

\begin{center}
\today
\end{center}


\definecolor{palegray}{gray}{0.95}
\begin{center}
\colorbox{palegray}{
  \begin{tabular}{rr}
  \begin{minipage}{0.1\textwidth}
    \includegraphics[width=22mm]{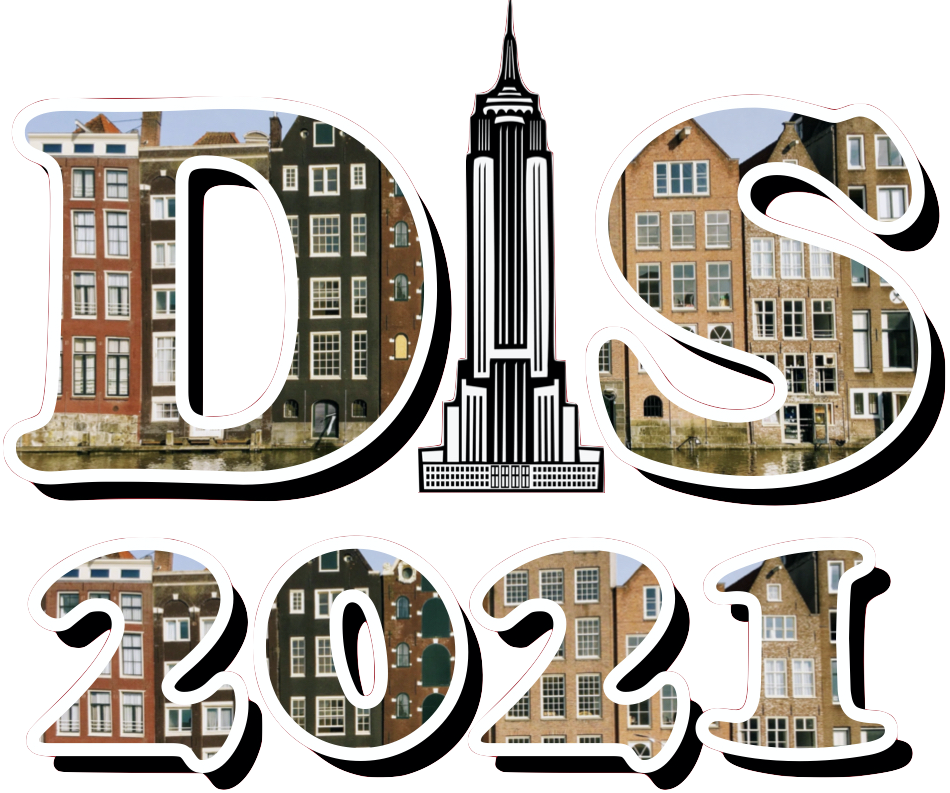}
  \end{minipage}
  &
  \begin{minipage}{0.75\textwidth}
    \begin{center}
    {\it Proceedings for the XXVIII International Workshop\\ on Deep-Inelastic Scattering and
Related Subjects,}\\
    {\it Stony Brook University, New York, USA, 12-16 April 2021} \\
    \doi{10.21468/SciPostPhysProc.?}\\
    \end{center}
  \end{minipage}
\end{tabular}
}
\end{center}

\section*{Abstract}
{\bf
We discuss two collider processes which combine a diffractively produced $\rho$ meson separated by a large rapidity gap from a hard exclusive scattering of a Pomeron on  a nucleon, giving rise to a lepton pair or to a second meson. These two processes probe the nucleon quark content described by generalized parton distributions in a very specific way.

}


\section{Introduction}
\label{sec:intro}
Diffractive events at high energy usually probe the low $x_\text{Bj}$ region, and consequently mostly the gluon content of the nucleon. A specific kinematical domain accessible at high energy electron-ion colliders allows, however, to access the valence quark region, opening a new experimental window on the nucleon tomography program. This domain is characterized by a large rapidity gap between a diffractively produced vector meson and a ``nucleon dissociation'' process reminiscent of deeply exclusive Compton scattering or deeply virtual meson  production.

In Refs.~\cite{Cosyn:2020kfe, Cosyn:2021dyr}, we analyzed diffractive reactions which allow access to certain combinations of generalized parton distributions (GPDs) that cannot be accessed in DVCS-like reactions, as for instance charge conjugation odd GPDs or chiral-odd quark GPDs. In these proceedings, we highlight results from these studies.

\begin{figure} [ht]
    \centering
    \includegraphics[width=0.8\textwidth]{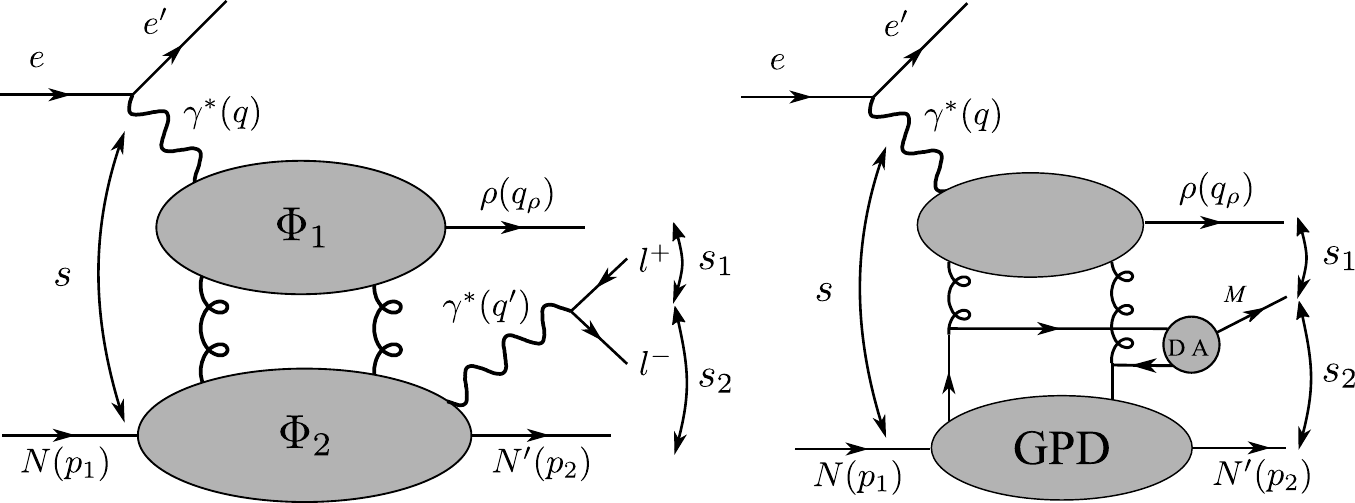}
    \caption{Left panel: the diffractive   amplitude (here for the $\rho$ + lepton pair) is written in the  $k_T$ -factorization approach as the convolution of two impact factors ($\Phi_1$,$\Phi_2$) and the Pomeron propagator, which at lowest order is a two gluon exchange. Right panel : the impact factor  $\Phi_2$ is then calculated (here for the $\rho$ + meson case) in collinear factorization, in terms of GPDs and DAs.  
    }
    \label{fig:process}
\end{figure}

\section{A hybrid factorized framework}

At large energy, the QCD amplitude may be written in the Regge-inspired $k_T$-factorization approach as the convolution of two impact factors $\Phi_1$ and $\Phi_2$ and the Pomeron ($\mathbb P$) propagator, which at lowest order is a  reggeized two gluon exchange and is subject to the BFKL evolution equations, see Fig.~\ref{fig:process}. In exclusive scattering, QCD collinear factorization enables to calculate the impact factors in terms of meson distribution amplitudes (DAs) and GPDs.

A characteristic feature of such a description  in the leading order approximation is that the cross sections are $s$-independent ($s$ being the initial $\gamma^{(*)} N$ squared invariant mass) which makes  their rates potentially quite large in the high energy domain.
Moreover, contrarily to the usual deeply virtual Compton scattering (DVCS) case, the skewness variable  is not related to the usual Bjorken variable $x_\text{Bj}$ but to a different (process-dependent) ratio which may be large even at high energy. Finally, the amplitudes turn out to depend on the so-called Efremov-Radyushkin-Brodsky-Lepage (ERBL) region of GPDs~\cite{Ivanov:2002jj,Pire:2019hos}  where GPDs are particularly unrestricted, in particular because the positivity constraints which relates them to usual quark distributions do not apply.

\section{Vector meson + M production}
\begin{figure}[ht]
\centering
\includegraphics[width=0.7\textwidth]{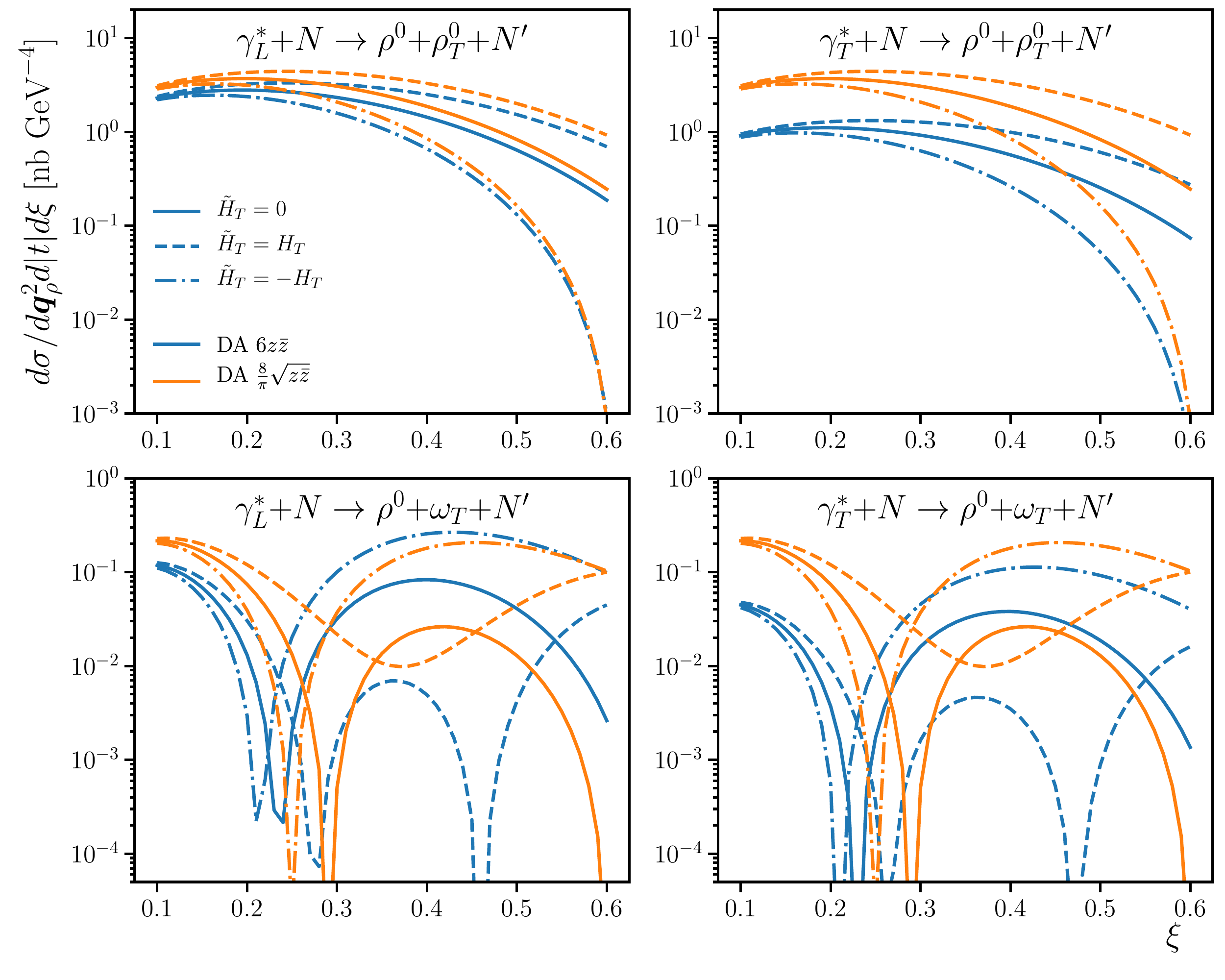}
\caption{$\xi$ dependence of the $\gamma^* + p \rightarrow \rho^0_L + (\rho^0_T/\omega_T) + p$ cross section for $Q^2=1 ~\text{GeV}^2$, $\Vec{q} _\rho^2=2 ~\text{GeV}^2$, $-t=(-t)_\text{min}$.  This process is sensitive to the chiral-odd nucleon GPDs.  We compare different GPD and DA models, see \cite{Cosyn:2020kfe} for details.}
\label{N_rhoT}
\end{figure}
\begin{figure}[ht]
    \centering
\includegraphics[width=0.6\textwidth]{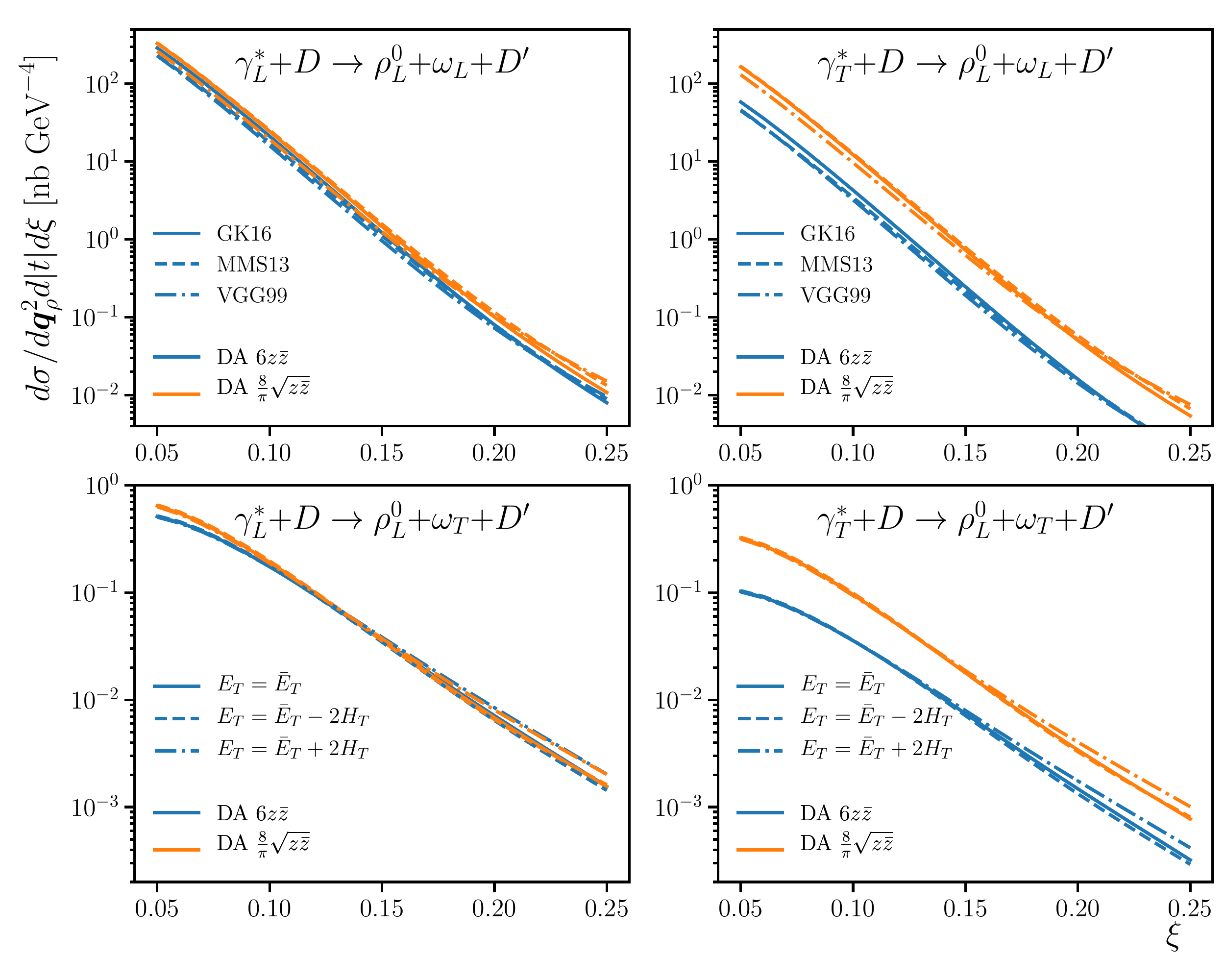}
\caption{$\xi$ dependence of the $\gamma^* + D \rightarrow \rho^0_L + \omega + D'$ coherent cross section for $Q^2=1 ~\text{GeV}^2$, $\bm q_\rho^2=2 ~\text{GeV}^2$.  This process is sensitive to the chiral-even deuteron GPDs ($\omega_L$ production, upper row) and chiral-odd deuteron GPDs ($\omega_T$ production, lower row).  We compare different GPD and DA models (see \cite{Cosyn:2020kfe} for details).}
    \label{fig:cross_xi_D}
\end{figure}

We follow the pioneering work \cite{Ivanov:2002jj}, where the exclusive process \begin{equation}
 e + N(p_1) \to e' + \rho_L^0(q_\rho) + M_2(p_{M_2}) + N'(p_2)
\end{equation}
 has been studied for the case $M_2 =\rho^+$ in the kinematical regime where the two $\rho$ mesons are separated by a large rapidity gap. The hard scale is the virtuality of the hard Pomeron, measured by the squared transverse momentum of the $\rho_L^0$-meson $\vec{q}_\rho^2 \approx (q-q_\rho)^2$\cite{Enberg:2006he}. 
 This hard scale ensures the small-sized configuration in the top part of the diagram ($\Phi_1$) and the separation of short-distance  and long-distance (GPD, DA) dynamics in the bottom part of the right panel of Fig.~\ref{fig:process}. The amplitudes  are particularly sensitive to yet quite unconstrained features of GPDs. In Ref.~\cite{Cosyn:2020kfe}, we revisited and enlarged the phenomenology of two-meson electroproduction in this hard diffractive regime, adding the coherent deuteron target case which has recently attracted much attention, and considering the $M_2=\pi,\omega$ channels. The skewness parameter $\xi$ is here given by
\begin{equation}
 \xi \approx \frac{s_1+Q^2}{2s-s_1+Q^2} \approx \frac{s_1}{2s-s_1}\,,
\label{s1}   
\end{equation}
where $s_1 = (q_\rho + p_{M_2})^2$ is the large invariant squared mass of the two produced mesons and $s=(q+p_{N_1})^2$ is the center-of-mass energy squared of the 
$\gamma^*p-$system, see Fig.~\ref{fig:process}.
Our results are  illustrated in Fig. \ref{N_rhoT} by the $\xi$ dependence of the $\gamma^* + p \rightarrow \rho^0_L + (\rho^0_T/\omega_T) + p$ cross section for $Q^2=1 ~\text{GeV}^2$, $\Vec{q} _\rho^2=2 ~\text{GeV}^2$ and the minimum momentum transfer to the nucleon $-t=(-t)_\text{min}$.  This process is sensitive to the more elusive chiral-odd nucleon GPDs and cross sections show large variations between the different considered DA and GPD inputs.

Fig.~\ref{fig:cross_xi_D} illustrates our results for the coherent deuteron reactions, where the $\rho^0_L \rho$ and $\rho^0_L \pi$ channels are forbidden by isospin constraints.  Here also, $\sigma_L$ cross-sections are larger than $\sigma_T$; cross sections are smaller for the deuteron case and drop faster with $\xi$, with the $\omega_L$ channels showing a larger drop. The deuteron GPDs \cite{Berger:2001zb,Cosyn:2018rdm} are calculated from the nucleon GPDs in a convolution model.

Our leading order analysis of the diffractive electroproduction of two mesons separated by a large rapidity gap  demonstrated that they are  a promising way to access nucleon and deuteron GPDs at EIC with a particular emphasis on some very bad known features of these non-perturbative objects. Both the chiral-even and chiral-odd GPDs are entering the amplitude at the leading twist level, their contributions being well separated in an angular analysis of the $\pi \pi$ (or $\pi \pi \pi$) decay products of the $\rho$ (or $\omega$) produced in the subprocess ${\mathbb P} h \to V_{L,T} h'$ where $h$ is a proton, neutron or deuteron, and $V_{L,T}$ the polarized vector meson.

\section{ Vector meson + lepton pair   production}

\begin{figure}[ht]
\centering
\includegraphics[width=0.9\textwidth]{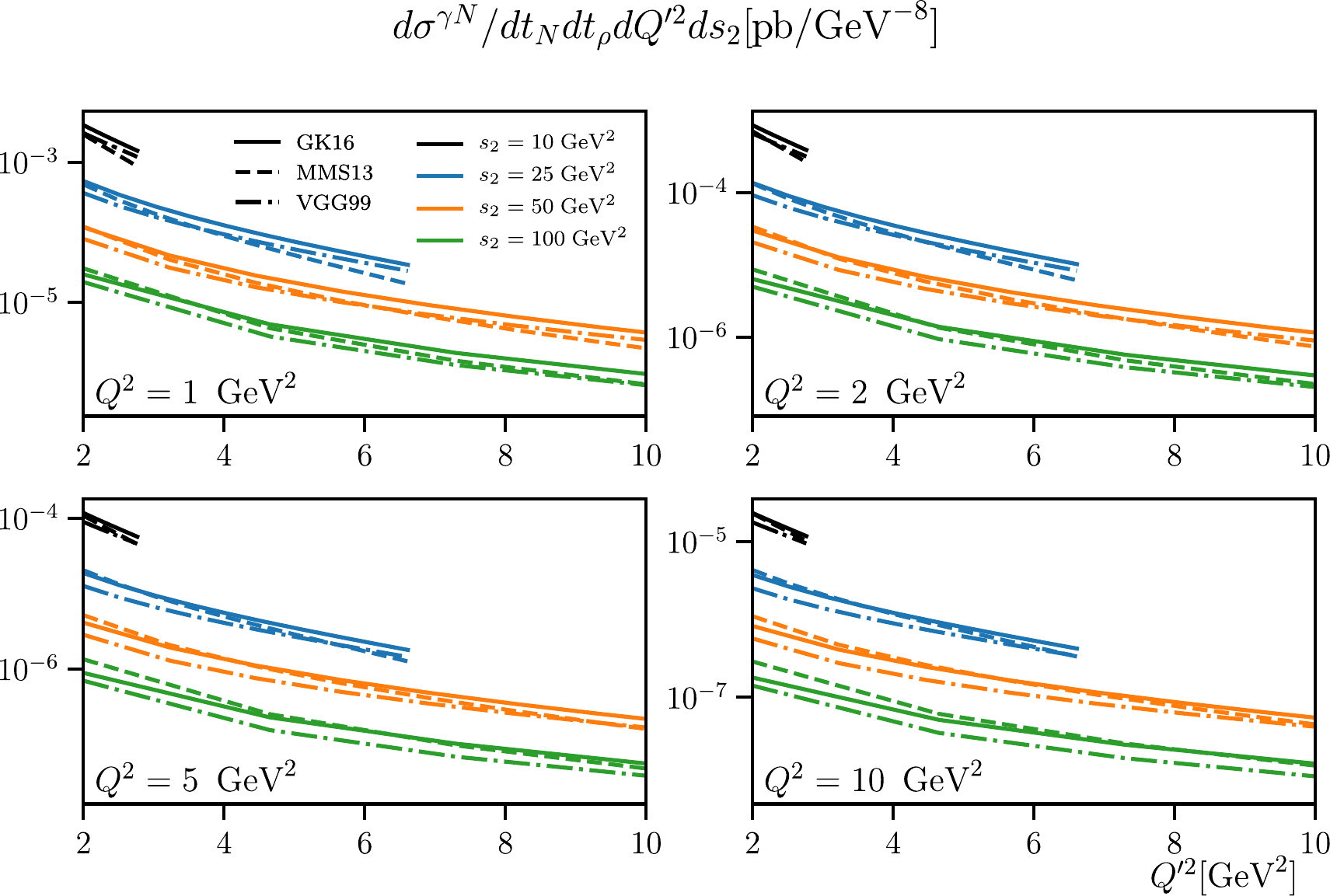}
\caption{Diffractive $\rho$ + dilepton photoproduction cross sections for different values of $Q^2$ at $t_N=-0.1~\text{GeV}^2$ and $t_\rho=t_{\rho}^{\text{min}}$, as a function of $Q'^2$ and for various values of $s_2$ (color), with three different GPD models (line style), see \cite{Cosyn:2021dyr} for details.  }
\label{diDVCS}
\end{figure}

This process, introduced in \cite{Pire:2019hos} and discussed in
\cite{Cosyn:2021dyr} is very reminiscent of the timelike Compton scattering $\gamma N \to \gamma^* N'$, with the Pomeron replacing the initial photon. To ensure a perturbative description of the exchanged Pomeron, one needs here to consider a virtual incoming photon and the amplitude of the subprocess 
\begin{equation}
    \gamma^*(q) + N(p_1) \to \rho(q_\rho) + \gamma^*(q') + N'(p_2)\,,
\end{equation}
now depends on two large scales $Q^2 = -q^2$ and $Q'^2 = q'^2$. The skewness parameter is 
\begin{equation}
 \xi \approx \frac{Q'^2}{2s_2 - Q'^2}    \,,
\end{equation}
with $s_2= (q'+p_2)^2$. Our studies, illustrated in Fig. \ref{diDVCS}, show that the cross sections are quite small at leading order, which can make a straightforward analysis of the process at the luminosities of planned electron-ion collider facilities quite difficult.  In terms of non-perturbative inputs, the calculations show much greater sensitivity to the nucleon GPD input than to the $\rho$ meson DA one~\cite{Cosyn:2021dyr}.  This GPD model sensitivity is due to the quite unique fact  that the amplitude only depends on their behaviour in the ERBL region, which is quite unrestricted by current data analysis of the DVCS process.  The cross section is dominated by the imaginary part of the Compton form factors~\cite{Pire:2019hos,Cosyn:2021dyr} and is maximized at small values of the hard scales $Q^2, Q'^2$ where higher order corrections to the formalism would be needed.

\section{Conclusion}
Data from a future high luminosity electron-ion collider will yield new opportunities to test
QCD techniques and to perform a complete tomography of nucleons.  The reactions presented here are just two of the possible processes that should be carefully analyzed and checked for their feasibility. The results present first steps for two reactions that allow to probe GPDs in regions or combinations where they are currently quite unconstrained. Much theoretical work remains to be done for these processes, in particular including higher order QCD corrections  and higher twist contributions.   

\section*{Acknowledgements}
The work of L.S. is supported by the grant 2019/33/B/ST2/02588 of the National Science Center in Poland. This project is also co-financed by the Polish-French collaboration agreements Polonium, by the Polish National Agency for Academic Exchange and COPIN-IN2P3 and by the European Union’s Horizon 2020 research and innovation programme under grant agreement No 824093.



\bibliography{biblio.bib}

\nolinenumbers

\end{document}